\documentclass[amsmath,amssymb,floatfix,showpacs,prb]{revtex4}

\usepackage{latexsym}

\usepackage{graphicx}
\usepackage{dcolumn}
\usepackage{bm}
\usepackage{color}

\newcommand{\ig}[2]{\includegraphics*[width=#1in]{#2}}
\newcommand{\ccirc}[1]{{\Large $\color[rgb]{#1,#1,#1} \bullet
\!\!\! \color{black} \circ$}}

\begin{document}

\preprint{}

\title{Atomistic Studies of Defect Nucleation during Nanoindentation
of Au (001)} 

\author{Anil Gannepalli}
\author{Surya K. Mallapragada}%
\email{suryakm@iastate.edu}
\affiliation{%
Department of Chemical Engineering, Iowa State University\\
Ames, Iowa 50011-2230
}%

\date{\today}

\begin{abstract}
Atomistic studies are carried out to investigate the formation and
evolution of defects during nanoindentation of a gold
crystal. The results in this theoretical study complement the
experimental investigations [J. D. Kiely and J. E. Houston,
Phys. Rev. B \textbf{57}, 12588 (1998)] extremely well. The defects
are produced by a three step mechanism involving nucleation, glide and
reaction of 
Shockley partials on the $\{111\}$ slip planes noncoplanar with the indented
surface. We have observed that slip is in the directions along which
the resolved shear stress has reached the critical value of
approximately 2 GPa. The first yield occurs when the shear 
stresses reach this critical value on all the $\{111\}$ planes
involved in the formation of the defect. The phenomenon of
strain hardening is observed due to the sessile stair-rods produced by
the zipping of the partials. The dislocation locks produced during the
second yield give rise to permanent deformation after retraction.
\end{abstract}

\pacs{62.20.Fe, 62.20.Qp}

\maketitle

\section{\label{sec:level1}Introduction}

Understanding the detailed mechanics of material deformation is a
fundamental challenge in materials science. In metals, the defect
structures produced during deformation influence the material
properties and behavior critically. \cite{mader63,whelan75}
The formation and evolution of such structures have their basis in
atomistic processes and the study of these nanoscale phenomena is
paramount to the understanding of macroscopic phenomena such as fracture,
friction, strain hardening and adhesion. The results of such research
will also greatly facilitate the design of novel materials with
desired properties. These insights into
material behavior can be exploited to create desired dislocation
patterns which can then be etched in a controlled manner 
to fabricate nanopatterns and nanostructures.\cite{wind01}

Nanoindentation experiments, with the advent of scanning probe
microscopes and advances in indentation techniques, are capable of
experimentally probing material properties and phenomena at the
nanoscale. \cite{corcoran97,kiely98,kiely99} At these atomic length scales,
the continuum models of deformation do not perform well and atomistic
methods need to be considered to investigate the nanoscale deformation
behavior. Advances in computational
capability and high performance techniques have enabled researchers to
investigate nanoindentation studies of comparable length
scales theoretically using molecular dynamics simulations.
\cite{kelchner98,zimmerman01} The
experiments, for most part, have emphasized quantitative
investigation of mechanical properties by measuring the
force displacement curves, and the theoretical computer simulations
have been targeted at studying the atomistic processes
involved in plastic deformation during indentation experiments. The
primary goal of such studies is to complement the experimental findings with
theoretical investigations in understanding the mechanisms of plastic
deformation in materials. 

In this paper, we present results of atomistic studies of
nanoindentation of a passivated gold surface. The objective of this work is
to study the atomistic processes responsible for plastic yield during
the initial stages of indentation and explain the experimentally
observed yield phenomena and defect structures.\cite{kiely98}

\section{Methodology}
\begin{figure}
\begin{minipage}[ht]{3.4 in}
	\ig{3.4}{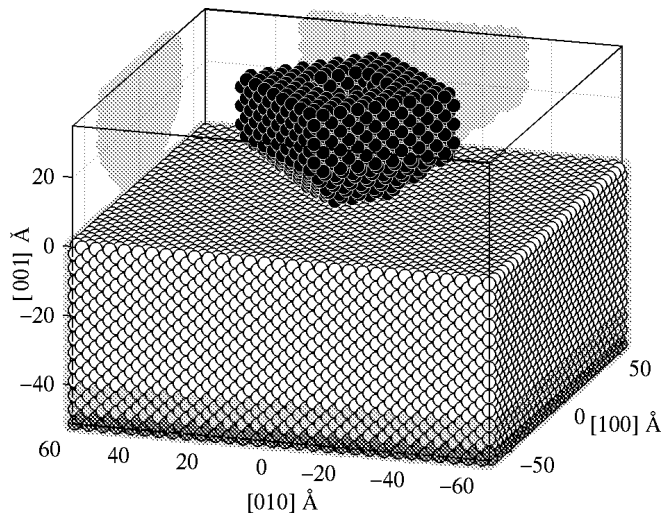}
	\caption{\label{fig:config} Atomic configuration of the
indenter and the gold substrate. \ccirc{0}~- indenter; \ccirc{1}
- dynamic gold substrate; \ccirc{0.75}~- temperature control region;
\ccirc{0.35}~- fully constrained boundary. The indenting face is a
square with edges along the $[110]$ and $[\bar{1}10]$ directions of
the gold crystal.} 
\end{minipage}
\end{figure}

The objective of this atomistic study is to investigate the defect 
nucleation during nanoindentation of a passivated Au(001) surface and
study the mechanisms leading to plastic deformation. The atomic
configuration of the system studied is illustrated in
Fig.~\ref{fig:config}. The gold substrate is 
modeled as a slab (122 \AA\ $\times$ 122 \AA\ $\times$ 50
\AA\ ) containing 46400 atoms with periodic boundary conditions
parallel to the surface. The orientation of the slab is such that the
directions [100], [010] and [001] coincide with x, y and z. 
The bottom layer is fully constrained and the
substrate size is sufficiently large to eliminate the finite size
effects. The indenter is an assemblage of atoms in diamondoid
cubic lattice arranged as a truncated pyramid with exposed $(111)$
facets and a 15 \AA\ $\times$ 15 \AA\ (001) square indenting face. The
indenter is oriented such that the edges of the indenting face are in
$[110]$ and $[\bar{1}10]$ directions with respect to the gold crystal.

We have employed the quantum Sutton-Chen (Q-SC) \cite{kimura98}
potential to model the gold atoms. This formulation includes the
quantum corrections to better predict mechanical properties, and
retains the simplicity of the original Sutton-Chen potential
\cite{sutton90} to facilitate the understanding of the underlying
physics of various processes. The indenter is modeled as a rigid body and the
indenter-surface interactions are purely repulsive,
$V(r)=\epsilon(r/\rho)^{-12}$ with $\epsilon$ = 25 meV and $\rho$ = 3
\AA\ , to eliminate the adhesive interactions and mimic the	
passivation of the gold surface in experiments.

We have used an extended version of the parallel MD package DL\_POLY
\cite{smith96} to perform the calculations. The dynamics of the substrate is
evaluated by integrating the Newtonian equations of motion using
Verlet-leapfrog method with a timestep of 1 fs. The gold substrate is
equilibrated to its minimum energy configuration at 300 K and the
indentation is simulated by advancing the indenter atoms by 0.0005 \AA\ at
every timestep, giving the indenter a velocity of 50 m/s. The temperature is
regulated by periodically scaling the velocities of the atoms of the deepest
non-constrained region of the substrate, away from the contact region
to minimize the interference of the temperature control mechanism with
the normal energy flow processes that occur in the contact region.

To understand the mechanics of plastic deformation during indentation,
atomic stress tensor \cite{egami82} $\bm{\sigma}$ is used to study the
distribution of stresses. The von Mises shear stress,
$\sqrt{J_2}$, proportional to the square root of the distortion energy,
is an indicator of the onset of plastic yielding \cite{dieter67} as
proposed by von Mises. The von Mises shear stress is given by the
square root of the second invariant of the deviatoric stress, $J_2$,
which is defined as

\begin{eqnarray}
J_2 = \scriptstyle\frac{1}{2} \displaystyle\mathrm{Tr} 
\left[
(\bm{\sigma} - p\bm{I})\cdot(\bm{\sigma}- p\bm{I})^{\mathrm{T}} 
\right],\\
p = -\scriptstyle\frac{1}{3} \displaystyle\mathrm{Tr}(\bm{\sigma}).
\end{eqnarray}

\noindent
where \textrm{Tr} denotes the trace of a matrix, $\bm{I}$ is the unit
matrix and $p$ is the local hydrostatic pressure.

In metals, plastic deformation occurs by the glide of dislocations on
the slip planes. In order to identify and characterize the
dislocations being nucleated during indentation we employ the slip 
vector analysis,\cite{zimmerman01} which provides information on the
Burgers vectors of dislocations. The slip vector is defined as,

\begin{equation}
\bm{s} = \frac{1}{n_s} \sum_\beta^{n_s}
\left(\bm{r}^{\beta}_t - \bm{r}^{\beta}_\theta \right).
\end{equation} 

\noindent
where, $n_s$ is the number of slipped neighbors $\beta$, of the
reference atom, and $\bm{r}^{\beta}_t$ and
$\bm{r}^{\beta}_\theta$ are the vector differences of atom
$\beta$ and the reference atom positions at
times $t$ and $\theta$, respectively. The slip vector given by the
above expression represents the Burgers vector of slip between the
plane containing atom $\alpha$ and its adjacent atomic
planes, in the time interval $[\theta,t]$. However, this is true only
in the case of single slip, where 
the reference atom is contained by only one slip plane. In the event
of multiple slip, where the atom is contained by two planes undergoing slip
simultaneously, the Burgers vector is different from the slip vector
given above. In any event, the slip vector will have a large magnitude
for inhomogeneous, non-affine deformation near the atom and can be
used to identify slipped regions.

The strains induced by indentation are studied by evaluating the
atomic strain tensor as formulated by Horstemeyer and
Baskes.\cite{horstemeyer00} This formulation is based on the
deformation gradient for a  material employing many-bodied
potential. The atomic Lagrangian Green strain tensor $\bm{E}$, used in
this study is given by,

\begin{subequations}
\begin{eqnarray}
\bm{E} = \frac{1}{2}\left(\bm{F}^\mathrm{T} \bm{F} - \bm{I} \right),\\
\bm{F} = \bm{X}\bm{Y}^{-1},\\
\bm{X} = \sum_\beta^m 
\left(\bm{r}^{\beta}_t\otimes\bm{r}^{\beta}_\theta\right),\\ 
\bm{Y} = \sum_\beta^m
\left(\bm{r}^{\beta}_\theta\otimes\bm{r}^{\beta}_\theta \right).
\end{eqnarray}
\end{subequations}

\noindent
where, $\bm{F}$ is the deformation gradient, $m$ is the number of
nearest neighbors $\beta$ of the reference atom,
$\bm{r}^\beta_\theta, \bm{r}^\beta_t$ have the same meaning as above
and $\otimes$ represents tensorial product. $\bm{E}$ will then
quantify the strain experienced by the reference atom in the time
interval $[\theta,t]$.

To investigate the mechanisms of dislocation nucleation and glide on
the slip planes, we study the resolved 
shear stresses on the slip planes along the Burgers vectors given by
the slip vector analysis. The resolved shear stress $\tau$ on a plane with
normal $\hat{\bm{n}}$ along the direction of slip $\hat{\bm{b}}$ is given by,

\begin{equation}
\tau_{(\bm{n})[\bm{b}]} = \hat{\bm{b}} \cdot \bm{\sigma} \cdot \hat{\bm{n}}
\end{equation}  

\noindent
Schmid law \cite{schmid50} states that a slip system is activated when
the resolved shear stress on that system reaches a critical value
called the critical resolved shear stress (CRSS).

\section{Results and Discussion}

\subsection{Indentation}

\begin{figure}
\begin{minipage}[ht]{3.4in}
\ig{3.4}{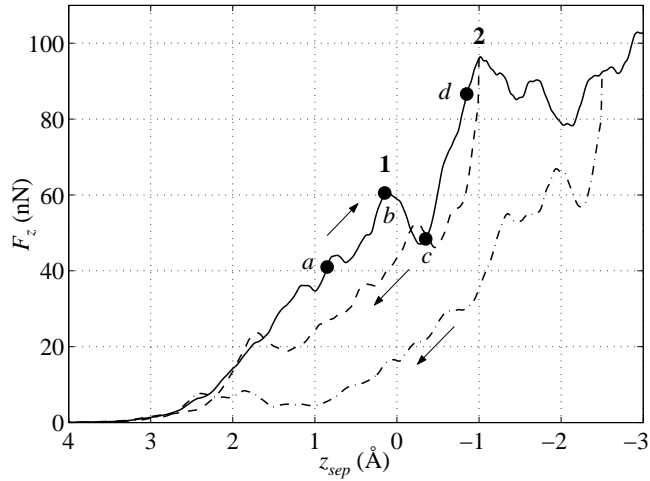}
\caption{\label{fig:F-z} Force versus distance curve during initial
stages of indentation. $(a)-(b)$ elastic response; $(b)$ onset of
first yield; $(b)-(c)$ first plastic yield event; $(c)-(d)$ second
elastic response at a higher force 
and with a higher slope indicating strain hardening. $\bm{1}$ and
$\bm{2}$ represent the first and second yield events.} 
\end{minipage}
\end{figure}

The force versus displacement curve for the initial stages of
indentation is shown in Fig.~\ref{fig:F-z}. The force $F_z$ is
calculated as the sum total of the forces exerted on the indenter atoms by
the substrate and the displacement $z_{sep}$ is the separation between
the indenter apex and the surface of the substrate before
indentation. Initially, the force curve displays elastic behavior
until the force decreases abruptly at the first yield point, marked as
($\bm{1}$) in Fig.~\ref{fig:F-z}. This
phenomenon is associated with the nucleation of a plastic event to
partially relieve the elastic stress in the contact region. This
observation is in excellent agreement with other theoretical
\cite{landman90,buldum98,gannepalli01,kelchner98,zimmerman01} and
experimental results.\cite{kiely98,kiely99,corcoran97} Upon further
indentation the force begins to rise again, displaying yet another
region of elastic behavior, until the substrate undergoes a second
yield event ($\bm{2}$) in Fig.~\ref{fig:F-z}. It is
interesting to see that the force curve has a higher slope in the
second elastic response region and the second yield occurs at a higher
force. This is indicative of strain hardening at the atomic scale
resulting in an increase in the yield modulus and strength.

\subsubsection{First Yield : Defect Nucleation}

\begin{figure*}
\begin{minipage}[ht]{7in}
	\ig{7}{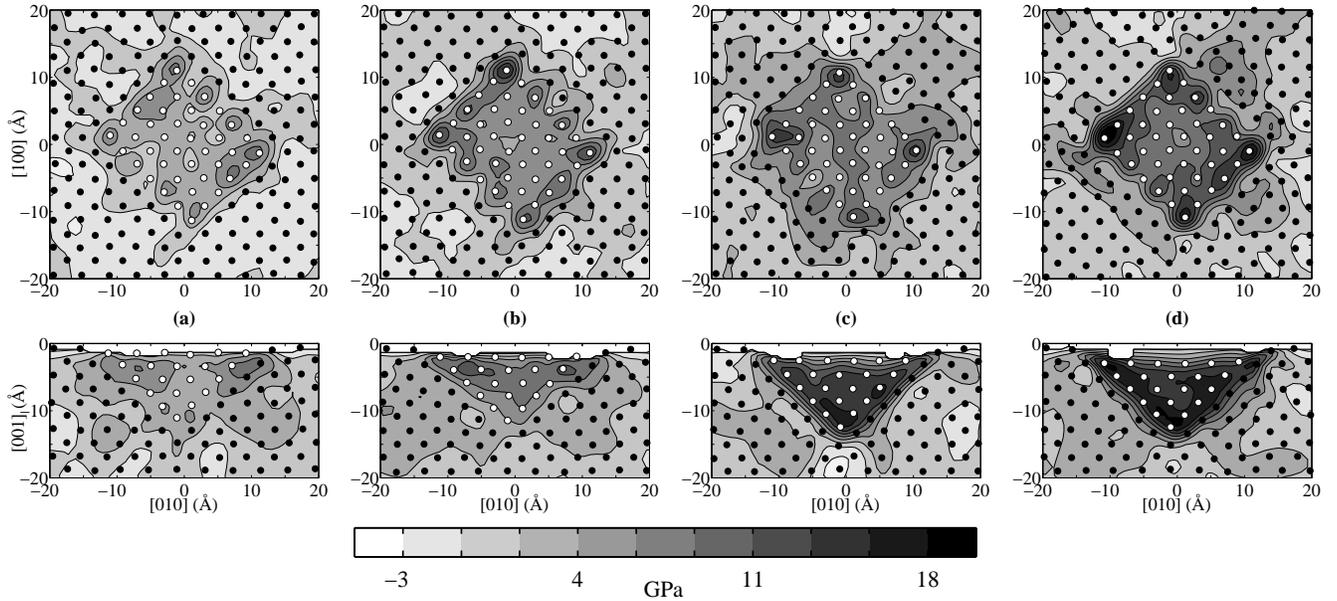}
	
	\caption{\label{fig:vonM} Contour plots of the atomic von
Mises shear stress  $\sqrt{J_2}$ in the indented region at four
stages of indentation marked $(a)-(d)$ in
Fig.~\ref{fig:F-z}. The contours are on [001] surface (upper row) just
beneath the indenter and [100] surface (lower row) at $x = 0$. 
\ccirc{1}~are the slipped atoms that comprise the  
defect nucleated during the first yield event and \ccirc{0}~are the
undeformed atoms. Stress is concentrated at the corners of the contact
region. Increase in $\sqrt{J_2}$ from $(a)$ to $(b)$ and $(c)$ to $(d)$
signifies  elastic responses and a drop from $(b)$ to $(c)$ indicates
plastic yield.}  

\end{minipage}
\end{figure*}

\begin{figure*}
\begin{minipage}[ht]{7in}
	\ig{7}{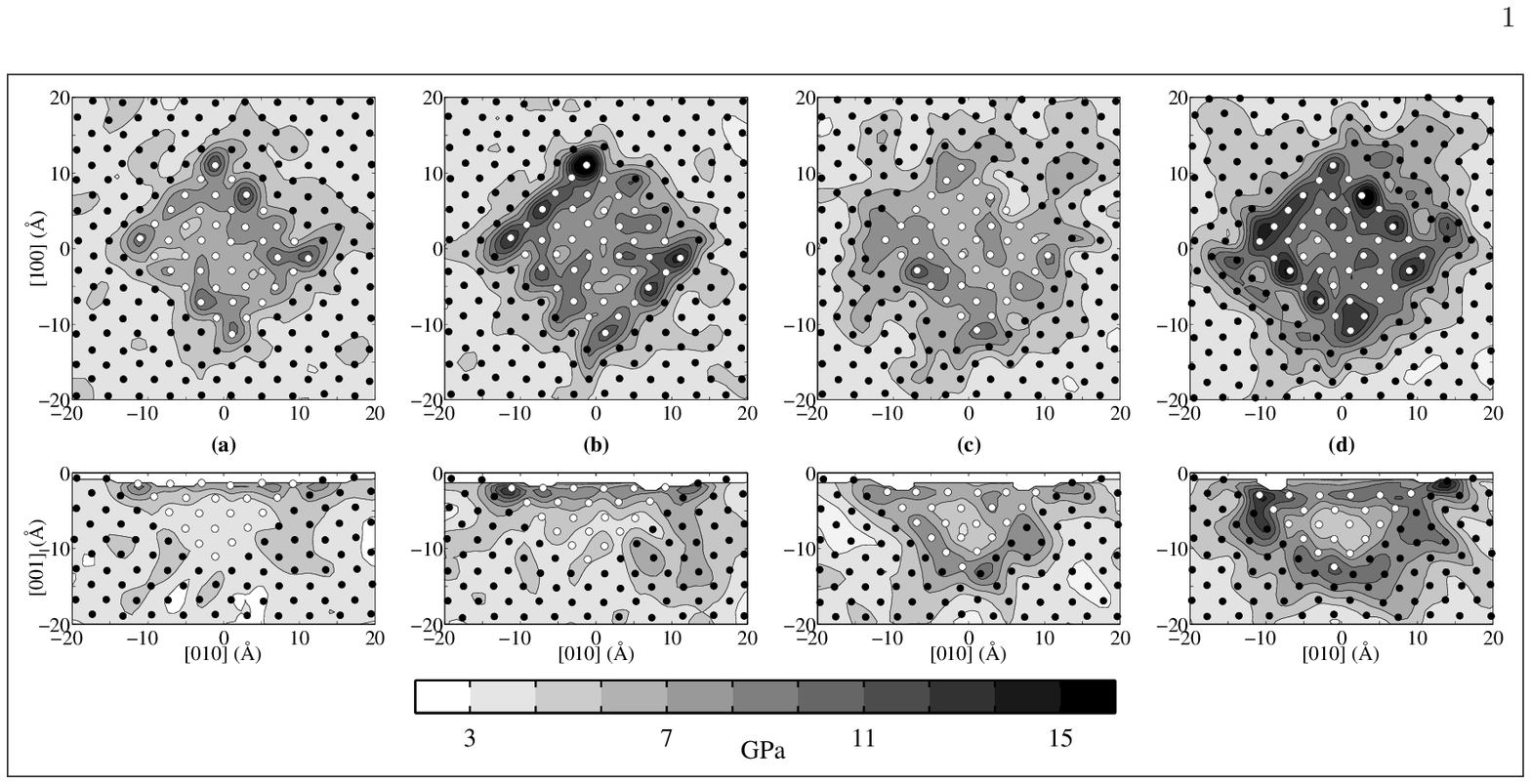}
	
	\caption{\label{fig:pres} Contour plots of the atomic hydrostatic
pressure  $p$ in the indented region at four stages of
indentation marked $(a)-(d)$ in Fig.~\ref{fig:F-z}. The contours are
on [001] surface (upper row) just beneath the indenter and [100]
surface (lower row) at $x = 0$. \ccirc{1}~are the 
slipped atoms that comprise the defect nucleated during the first
yield event and \ccirc{0}~are the undeformed atoms. Stress is
concentrated at the corners of the contact region. After the first
yield $(c)$, a compressive strain of 0.052 in the defect gives rise to
increased pressure of the order of $10-15$ GPa.}  
\end{minipage}
\end{figure*}

To gain insight into this behavior, the evolution of the stress
profiles in the contact region during indentation are
analyzed. Figures \ref{fig:vonM} and \ref{fig:pres}
show the von Mises shear stress $\sqrt{J_2}$ and hydrostatic
pressure $p$ profiles in the region directly beneath the indenter at
various stages of indentation marked $(a)-(d)$ in
Fig.~\ref{fig:F-z}. Figures \ref{fig:vonM}(a), (b) and
\ref{fig:pres}(a), (b) show that as the 
indentation proceeds from $(a)$ to $(b)$, an increase in $\sqrt{J_2}$, a
measure of the elastic stored energy, substantiates the elastic
response seen in the force curve in this regime. At point $(b)$
the elastic stress reach a threshold beyond which plastic
deformation occurs that partially relieves and
dissipates the elastic energy from the surface as seen in
 \ref{fig:vonM}(c). This behavior is in
accordance with the von Mises criterion,\cite{dieter67} which
suggests a critical value for $\sqrt{J_2}$ for the onset of plastic
activity.  Upon further indentation from $(c)$ to $(d)$
$\sqrt{J_2}$ increases again implying another elastic response regime,
which culminates in the second yield event. 

\begin{figure*}
\begin{minipage}[ht]{7in}
	\ig{7}{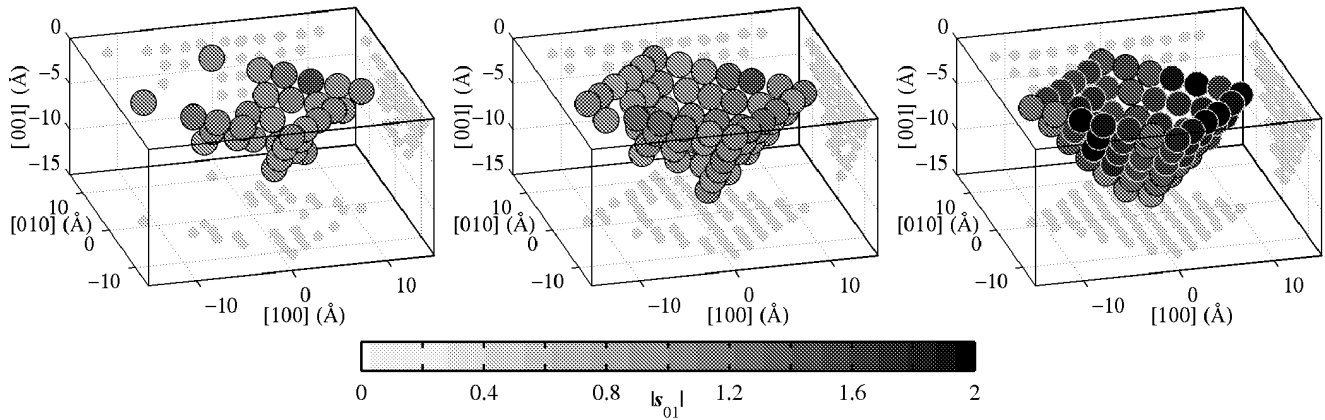}
	
	\caption{\label{fig:defect} Snapshots of deformed region
depicting the evolution of the dislocation structures nucleated during the
first yield event ($(b)-(c)$ in Fig.~\ref{fig:F-z}). Greyscale
represents $|\bm{s}_{01}|$.} 
\end{minipage}
\end{figure*}

\begin{figure*}
\begin{minipage}[ht]{7in}
	\ig{7}{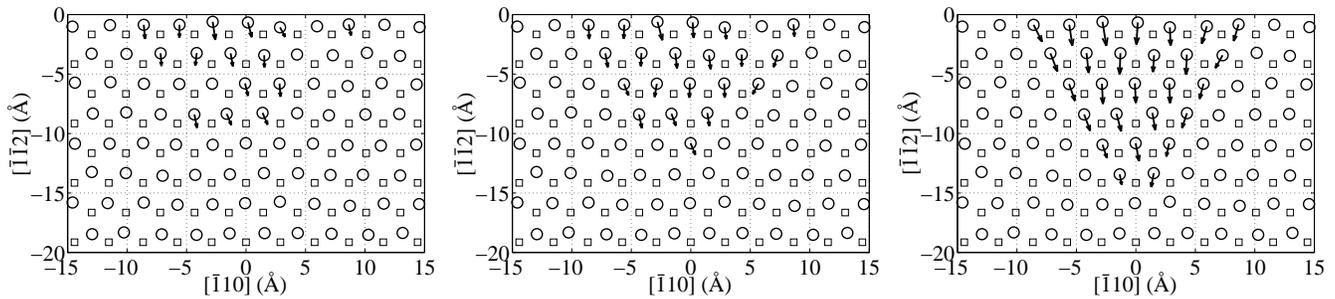}

	\caption{\label{fig:svec} Slip vector ($\bm{s}_{01}$) maps on
$(111)$ plane corresponding to the snapshots in
Fig.~\ref{fig:defect}. \ccirc{1}~represent the atoms of the slipped
plane and $\square$ represent the atoms of the unslipped plane
adjacent to the slipped region.} 
\end{minipage}
\end{figure*}

To study the nature of plastic deformation and characterize the defect
structures nucleated, the deformed regions are identified by
the slip vector $\bm{s}_{01}$, where $\bm{0}$ represents
the initial undeformed state and $\bm{1}$ represents the state after
the first yield event.
Three snapshots of the deformed region at various stages of defect
nucleation between $(b)$ and $(c)$ are shown in Fig.~\ref{fig:defect}
to illustrate the evolution of the defect structure. From the slipped
atoms shown in Fig.~\ref{fig:defect} it is seen that dislocation loops
nucleate on the four $\{111\}$ planes at the surface and extend
into the solid. These dislocation loops grow in size and intersect
with the loops on the adjacent planes forming a pyramidal defect
structure as seen in Fig.~\ref{fig:defect}(c).
Figure \ref{fig:svec} shows the corresponding slip
vectors of the atoms on one of the slip planes, $(111)$. 
From Fig.~\ref{fig:svec}(c) the
magnitude of the slip vector of the atoms on the $(111)$
plane is close to 1.66 \AA\ along $[11\bar{2}]$, which is consistent
with the $\langle112\rangle$ partial dislocations on \{111\} planes in
gold. The dislocation nucleated on the $(111)$ plane is therefore
the $\frac{1}{6}[11\bar{2}]$ Shockley partial, and similarly
$\frac{1}{6}[\bar{1}1\bar{2}]$, $\frac{1}{6}[1\bar{1}\bar{2}]$ and
$\frac{1}{6}[\bar{1}\bar{1}\bar{2}]$, partials are nucleated on 
$(\bar{1}11)$, $(1\bar{1}1)$ and $(\bar{1}\bar{1}\bar{1})$ planes,
respectively. Thus the defect consists of intersecting intrinsic stacking
faults on the four $\{111\}$ planes, which intersect the (001) surface
with four fold symmetry. This pyramidal defect structure is in
excellent agreement with the experimentally observed permanent
deformation structures. \cite{kiely98} 

From continuum elastic theory \cite{johnson85} of indentation by a
rigid flat frictionless punch, similar to the atomic indenter used in
this study, the stresses reach a theoretically infinite value at the
edges of the indenter. This observation, at the atomic scale, is
validated by the large concentrations of stresses at the periphery
of the contact region as seen in Figs.~\ref{fig:vonM}(b) and
\ref{fig:pres}(b), in agreement with previous
studies.\cite{landman90,gannepalli01}  
Surface nucleation of partial dislocations in areas of enhanced
stress is well supported experimentally
\cite{ning96,doerschel94} and theoretical models of
surface dislocation nucleation at stress concentrators
\cite{kamat90,gao94,beltz95,zou96} are well established. The corners
of the contact region act as stress concentrators and serve as the
sources of nucleation of Shockley partials on the surface. 
Under the influence of the stresses, these partials glide on the
\{111\} planes, the dominant slip planes in gold, forming intrinsic stacking
faults as seen in Fig.~\ref{fig:svec}.
This slip results in the flow of part of the elastic energy from the
contact surface to the sheared surfaces and is seen as an increase in
$\sqrt{J_2}$ in the region of slip in Fig.~\ref{fig:vonM}(c).
The strain undergone by the atoms in the deformed region during defect
nucleation is quantified by the mean Lagrangian strain tensor in Voigt
notation, $\bm{E} = \begin{bmatrix} 0.068 &
0.064 & -0.166 & 0.007 & -0.002 & 0.02 \end{bmatrix}$. Thus the
observed strain is in $\langle100\rangle$ directions and gives a
volumetric strain of -0.052 in the pyramidal defect, giving rise to
higher pressures of the order of $10-15$ GPa (with a bulk modulus of
207 GPa) above the ambient pressures as seen in Fig.~\ref{fig:pres}(c).

\begin{figure}
\begin{minipage}[ht]{3.4 in}
\ig{3.4}{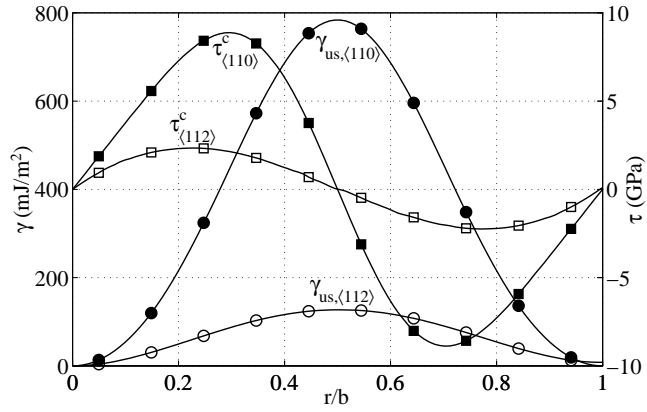}
\caption{\label{fig:gsf} Generalized stacking fault energies $\gamma$
(circles) and theoretical shear stresses $\tau^{th}$ (squares) for
slip systems $\{111\}\langle112\rangle$ (open) and
$\{111\}\langle110\rangle$ (solid). Lower $\tau^c_{\langle112\rangle}$
indicates a more energetically favorable $\{111\}\langle112\rangle$
slip system.} 
\end{minipage}
\end{figure}

\begin{figure*}
\begin{minipage}[ht]{7in}
	\ig{7}{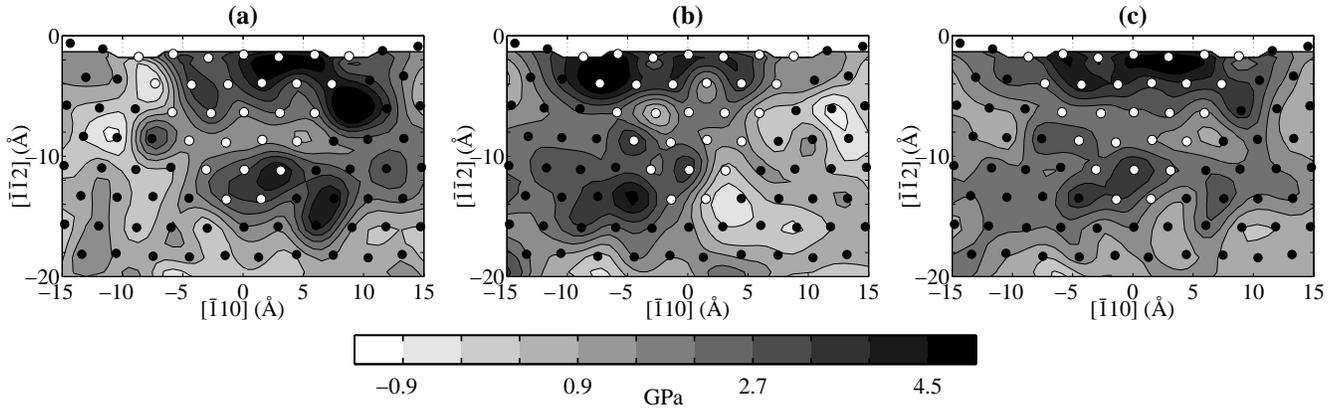}
	
	\caption{\label{fig:rss1} Resolved shear stresses on $(111)$
plane, just before the first yield point, along favored directions of
slip: (a) $[01\bar{1}]$; (b) $[10\bar{1}]$; (c)
$[11\bar{2}]$. \ccirc{1}~are the slipped atoms that constitute the stacking
fault and \ccirc{0}~are the atoms in the undeformed
region. In some regions $\tau_{(111)[11\bar{2}]}$ is smaller than
$\tau_{(111)[10\bar{1}]}$ and $\tau_{(111)[01\bar{1}]}$.}   
\end{minipage}
\end{figure*}

In order to understand why this particular slip system
$\{111\}\langle11\bar{2}\rangle$ has been activated, we study the
resolved shear stresses (RSS), as the direction of slip has been
associated with maximum RSS. Fig.~\ref{fig:rss1} shows the
distribution of $\tau_{(111)[10\bar{1}]}$,
$\tau_{(111)[01\bar{1}]}$ and $\tau_{(111)[11\bar{2}]}$ on the $(111)$
slip plane along $[10\bar{1}]$, $[01\bar{1}]$ and $[11\bar{2}]$,
respectively. These are the favored slip directions \cite{schmidfac}
on $(111)$ plane for the current stress state $(\bm{\sigma} \approx
\sigma_{zz})$. It is seen
that even though $\tau_{(111)[10\bar{1}]}$ and
$\tau_{(111)[01\bar{1}]}$ have higher concentrations near the surface than 
$\tau_{(111)[11\bar{2}]}$, the slip occurs along $[11\bar{2}]$. This
incongruency of slip occurring in a lower RSS direction is consistent
with the findings of other researchers \cite{zimmerman01} and can be
explained by the concept of generalized stacking fault energy (GSF)
$\gamma$, introduced by Vitek.\cite{vitek68,vitek72} GSF is the
energy per unit area of a fault plane created by the rigid slip of one
half of a perfect lattice relative to the other along a slip
plane in a general slip direction. Fig.~\ref{fig:gsf} shows the
unrelaxed GSF in the dominant slip directions $[110]$ and $[112]$. 

The maximum value of $\gamma$ in the
direction of slip, called the unstable stacking energy
$\gamma_{us}$,\cite{rice92} 
is the energy barrier to be overcome during slip. It is seen from
Fig.~\ref{fig:gsf} that $\gamma_{us}$ displays a strong directional
dependence and slip along $[112]$ has a lower energy barrier than
$[110]$ and is thus more favorable as observed in this study. However,
$\gamma$ is a static quantity and is not an appropriate measure to
describe the dynamics of slip. A better quantity would be the
theoretical shear stress required to initiate and maintain the
slip along the slip direction. This shear stress is given by

\begin{equation}
\tau_{\{\bm{n}\} \langle\bm{r}\rangle}^{th} = \frac{\partial
	\gamma_{\bm{n}}}{\partial \bm{r}}
\end{equation}

\noindent
where, $\gamma_{\bm{n}}$ is the GSF on a slip plane with normal
$\bm{n}$ and $\bm{r}$ is the displacement vector. A plot of this stress for
slip on $\{111\}$ plane along $\langle110\rangle$ and $\langle112\rangle$
directions is shown in Fig.~\ref{fig:gsf}. The maximum value of
$\tau^{th}$ is the resolved shear stress that is required to complete
the slip along the particular direction. This is the critical resolved
shear stress (CRSS) $\tau^c$ referred to in Schmid law
\cite{schmid50} as stated above. From Figs.~\ref{fig:gsf} and
\ref{fig:rss1} $\tau^c_{\{111\}\langle112\rangle}$ is 2.34 GPa and is
smaller than the observed $\tau_{(111)[11\bar{2}]}$ values. On the
other hand $\tau^c_{\{111\}\langle110\rangle}$ has a value of 8.88 GPa
and is much higher than the observed $\tau_{(111)[11\bar{2}]}$
values. Thus, the observed slip direction is $[11\bar{2}]$ rather than
$[01\bar{1}]$ or $[10\bar{1}]$. Second and higher derivatives of
$\gamma$ could be used to further refine the dynamics of slip, but it
is beyond the scope of this paper and for the present study $\tau$
would suffice.  

\begin{figure}
\begin{minipage}[ht]{3.4 in}
	\ig{3.4}{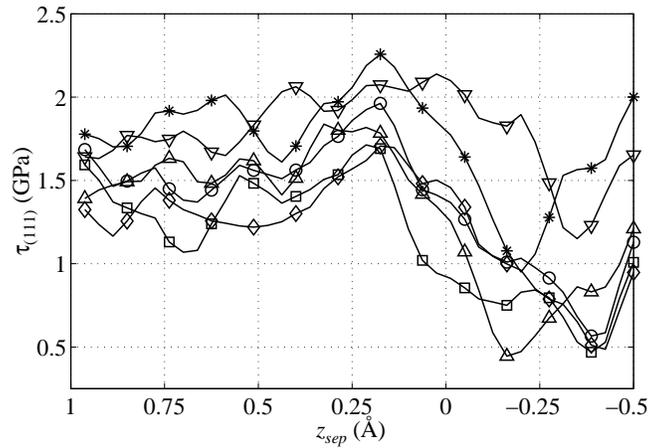}
	\caption{\label{fig:mrss} Mean resolved shear stresses in the
deformed region for slip systems: $\Box~(111)[01\bar{1}$;
$\Diamond~(111)[10\bar{1}]$; \ccirc{1}~$(111)[11\bar{2}]$; {\Large
$\ast$}~$(\bar{1}11)[\bar{1}1\bar{2}]$; {\small
$\bigtriangledown$}~$(1\bar{1}1)[1\bar{1}\bar{2}]$; {\small
$\bigtriangleup$}~$(\bar{1}\bar{1}1)[\bar{1}\bar{1}\bar{2}]$. The
resolved shear stresses reach a threshold at the yield point. The
maximum resolved shear stress is in the range of $1.8-2.3$ GPa.}
\end{minipage}
\end{figure}

The smallest of the directionally dependent $\tau^c$ represents the
ideal shear strength of the crystal is 2.34 GPa in excellent
agreement with the experimental estimates of 1.5 to 2.0
GPa.\cite{kiely98,corcoran97} The high $\tau$ values of 5 GPa, in
Fig.~\ref{fig:rss1}, greater than the theoretical estimate of the ideal
shear strength might seem out of order, but it needs to be
clarified that the theoretical value is based on a block like
shear and such instantaneous rigid slip cannot be expected during the
actual nucleation and propagation of slip. These $\tau$ values are
also quite high compared to the experimental CRSS because the values
in Fig.~\ref{fig:rss1} are highly localized and are inaccessible to
experimental investigations. The shear stress values deduced from
experiments represent the mean value of the shear stresses in the
region local to indentation and a plot of such a mean of $\tau$ is
shown in Fig.~\ref{fig:mrss}. It can be seen that these values reach a
maximum and drop abruptly at the first yield point validating the
manifestation of Schmid law at the atomic scale. The maximum RSS on
$(111)$ is along $[11\bar{2}]$ and reaches a value of 1.95 GPa and
on the other $\{111\}$ planes the maximum RSS are in the range of 1.8
to 2.3 GPa. These values agree exceptionally well with the
experimental estimates of 2 GPa.

The plastic strains produced by indentation are complex and activation
of multiple slip systems is necessary to accommodate these general
yields.\cite{vonMises28} Groves and Kelly \cite{groves63} have
predicted the active slip systems by calculating the strain produced
by a given slip system \cite{bishop53} and identifying the
systems that contribute to the observed strain. Such a geometrical
analysis \cite{reid73} for compression in $[001]$, which is the
observed stress state just before the first yield point, predicts
activation of slip on the four $\{111\}$ planes resulting in plastic
strains in $\langle100\rangle$. These predicted slip planes and
strains are identical to those observed at the atomic scale in this
study.

In the event of the presence of multiple sets of independent slip
systems capable of producing the required strain, as is the case with
f.c.c. crystals, Bishop and Hill \cite{bishop51a,bishop51b} proposed a stress
criterion for yielding that requires the attainment of CRSS on the
active slip systems, without exceeding CRSS on the inactive
systems. The observed slip directions conform to the above criterion
with the shear stresses reaching their critical values in
$\langle112\rangle$, but not in $\langle110\rangle$.

\subsubsection{Defect Nucleation Mechanism}

\begin{figure*}
\begin{minipage}[ht]{7 in}
	\ig{2.3}{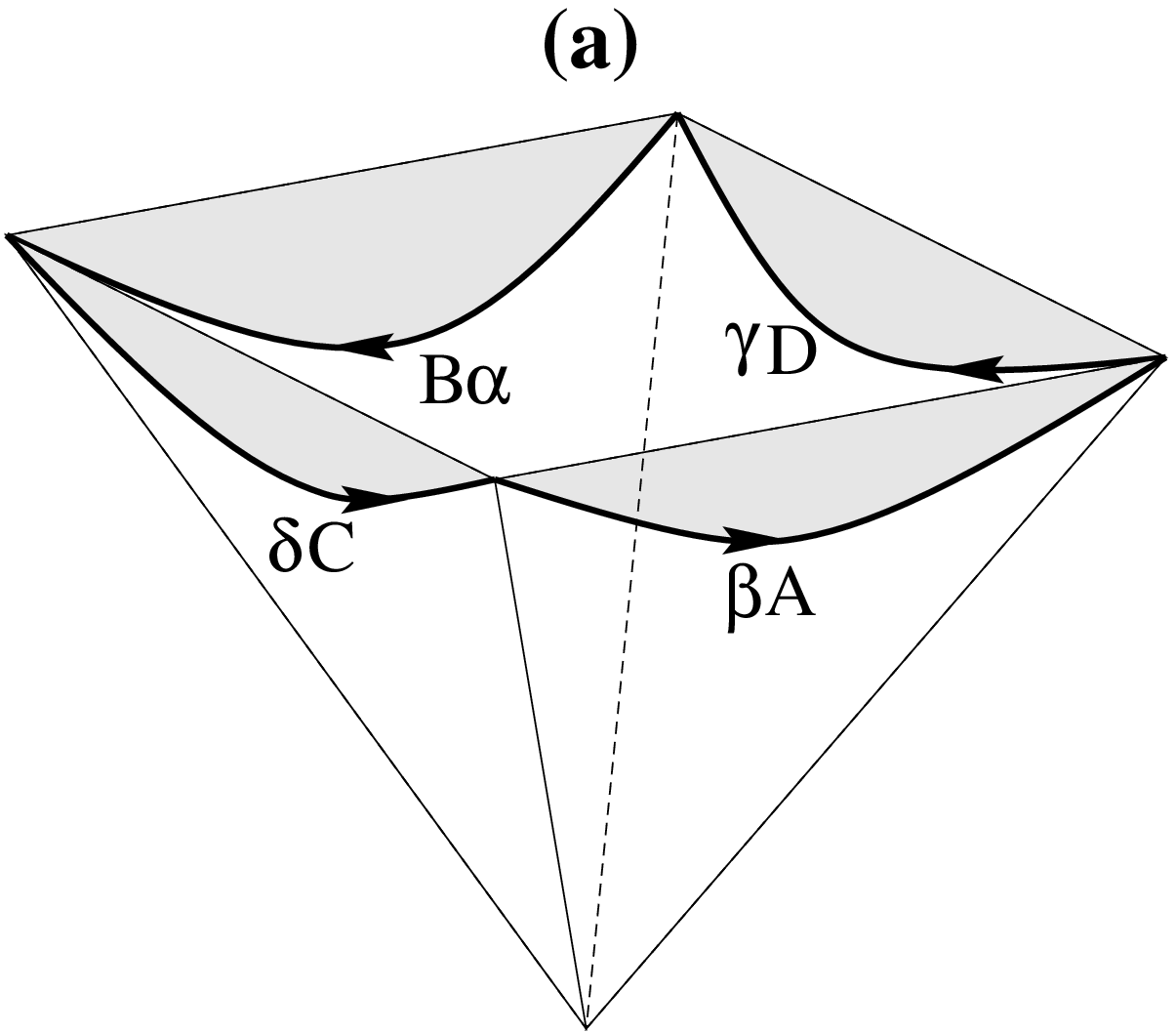}
	\ig{2.3}{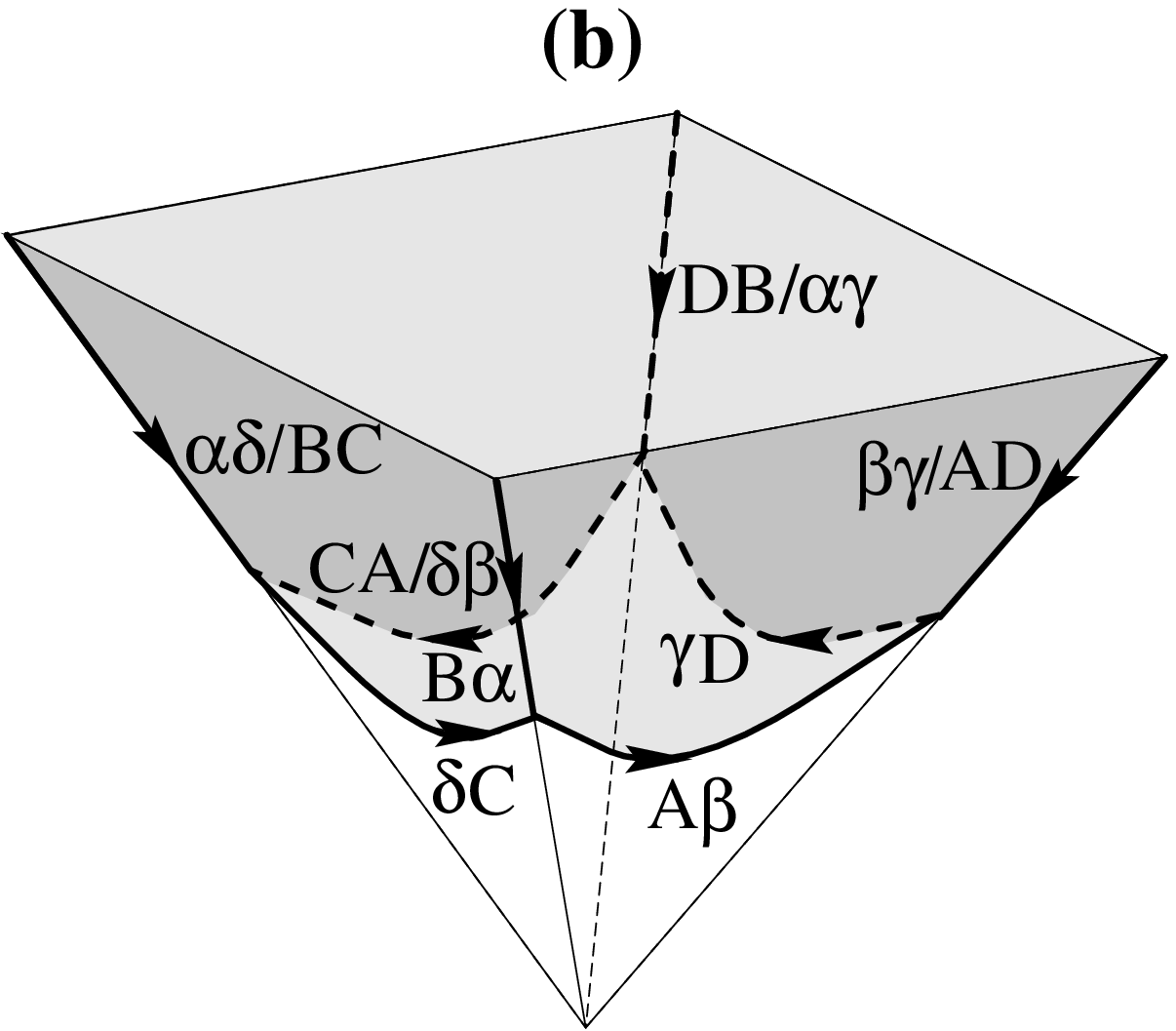}
	\ig{2.3}{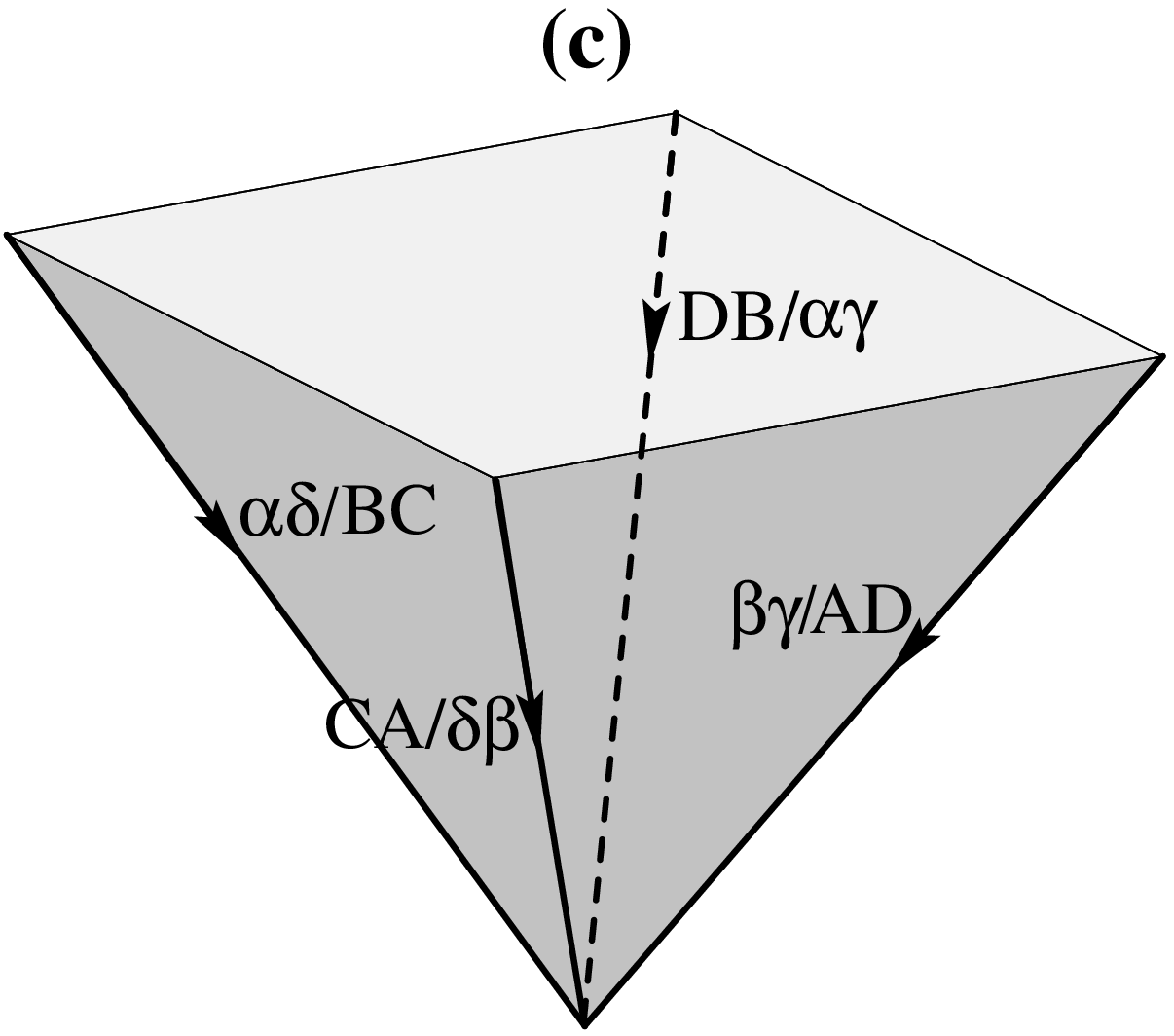}
	\caption{\label{fig:schm} Various stages in the formation of
the pyramidal defect structure. Shaded regions represent intrinsic
stacking faults. (a) nucleation and glide of partials; (b)
dislocation loop growth and zipping resulting in sessile stair-rods;
(c) pyramidal defect structure.} 
\end{minipage}
\end{figure*}

Based on the results and discussion presented above, we propose a
three step mechanism for the formation of the pyramidal defect during
indentation of Au (001). It is convenient 
for the following discussion to use Thompson's notation for Burgers
vectors and planes and refer to Fig.~\ref{fig:schm} for an
illustration of the mechanism.

\begin{trivlist}
\item[] \textit{Dislocation nucleation}: Surface indentation of Au (001)
with an indenter results in large concentration of stresses at the
corners of the contact region. These stress concentrators, where the
RSS on the $\{111\}$ planes reach the CRSS, act as the
sources for surface nucleation of Shockley partials
($\frac{1}{6}\langle 11\bar{2} \rangle$)
B$\alpha$, A$\beta$, $\gamma$D and $\delta$C on the slip planes
$\mathrm{(\bar{a}),(\bar{b}),(c)}$ and $\mathrm{(d)}$ respectively. 
\item[] \textit{Dislocation glide}: These partials, under the influence of the
external stress due to indentation, glide away from the surface forming
intrinsic faults on the slip planes. 
\item[] \textit{Dislocation reaction}: As the dislocation loops grow,
the partials attract each other in pairs and zip to form sessile stair-rods
along AC, AD, BD and BC according to the following reactions:

\begin{eqnarray}
\delta\mathrm{C} + \beta\mathrm{A} = \delta\beta/\mathrm{CA} \nonumber \\
\mathrm{A}\beta + \mathrm{D}\gamma = \mathrm{AD}/\beta\gamma \nonumber \\
\gamma\mathrm{D} + \alpha\mathrm{B} = \gamma\alpha/\mathrm{DB}
\nonumber \\
\mathrm{B}\alpha + \mathrm{C}\delta = \mathrm{BC}/\alpha\delta
\end{eqnarray}

\noindent
In vector notation the energetically favorable reactions are of the type

\begin{equation}
\scriptstyle\frac{1}{6}\displaystyle[11\bar{2}] -
\scriptstyle\frac{1}{6}\displaystyle[\bar{1}1\bar{2}] =
\scriptstyle\frac{1}{3}\displaystyle[100] 
\end{equation}

\end{trivlist} 

The final defect, therefore, consists of a pyramid of intrinsic stacking
faults on \{111\} planes, which intersect the (001) surface with a
four fold symmetry, and the $\langle01\bar{1}\rangle$ edges of the
pyramid consist of low energy sessile stair-rod dislocations. These
sessile stair-rods act as barriers to further glide giving rise to the
observed strain hardening during indentation beyond the first yield.

To examine the dependence of the defect structures on the orientation of
the indenter with respect to the crystallographic axes of the gold
substrate, the calculations were repeated
with the edges of the indenting face in (100) and (010)
directions. The defect produced was similar to the pyramidal structure
seen above where the high stresses at the corners of contact region
nucleate partials on the four $\{111\}$ planes. From the physics of
the defect nucleation presented above, the above mechanism can be
generalized to indentation of other $\{111\}$ and $\{110\}$ surfaces as
well. It 
can be deduced that the indentation of $\{111\}$ surface will produce a
tetrahedral defect structure displaying the threefold symmetry
observed in experiments. \cite{kiely98} And similarly the hexagonal
nature of the indent produced by indenting the $\{110\}$ surface is
consistent with the proposed mechanism. Thus, the geometry of
the defect structures is independent of the indenter orientation, but
is a characteristic of the crystallography of the surface of the indented
crystal.

\subsubsection{Second Yield : Dislocation Locks}

\begin{figure}
\begin{minipage}[ht]{3.4 in}
	\ig{3.4}{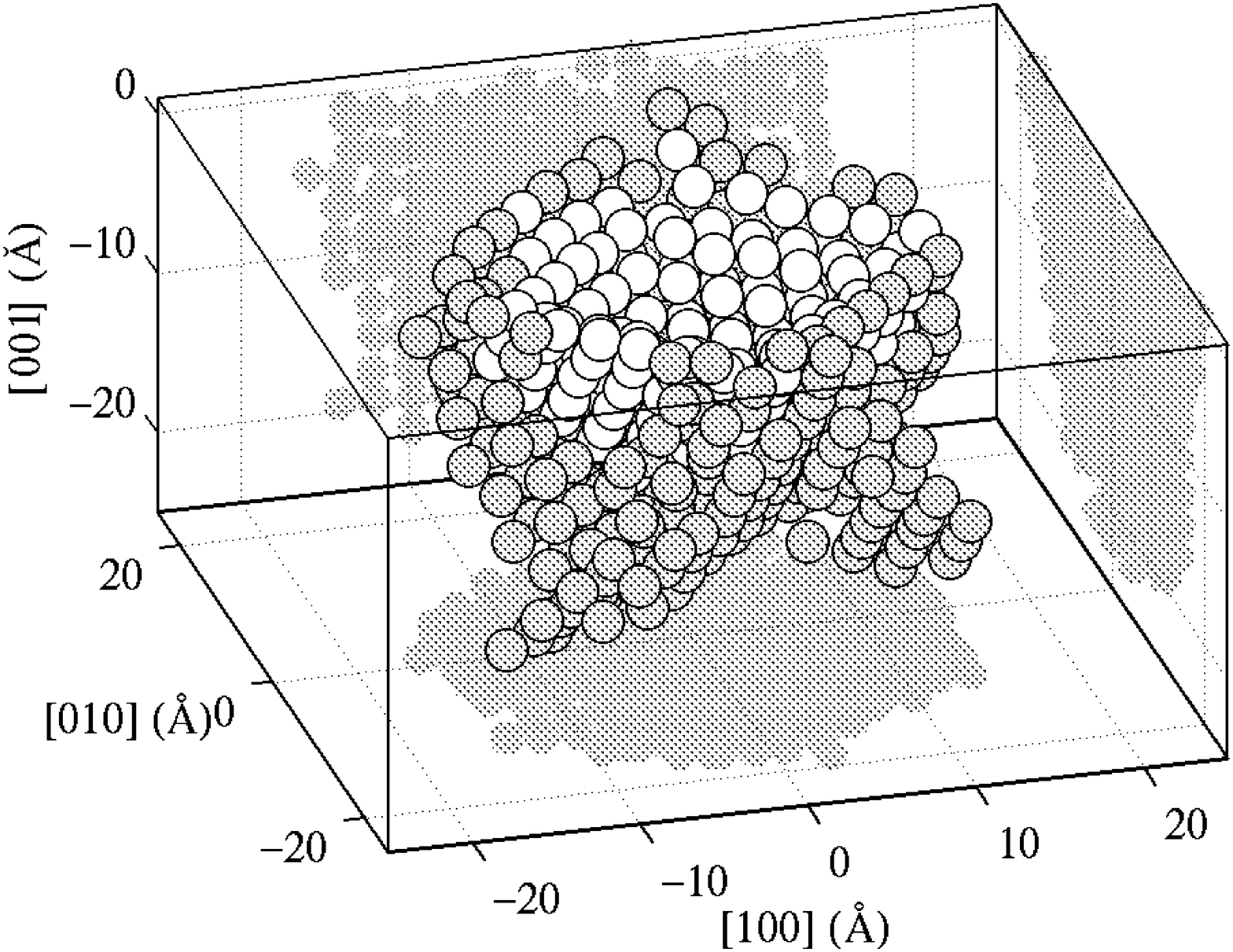}
	\caption{\label{fig:lock} Deformed region illustrating the
dislocation locks formed during the second yield. \ccirc{1}~represent
the atoms that slipped during the first yield ($|\bm{s}_{01}| > 0$) and
\ccirc{0.75}~are the atoms that underwent slip during the second yield
event ($|\bm{s}_{12}| > 0$). Dislocation loops extend beyond the
adjacent faces of defect structure forming dislocation locks.}   
\end{minipage}
\end{figure}

\begin{figure}
\begin{minipage}[ht]{3.4 in}
	\ig{3.4}{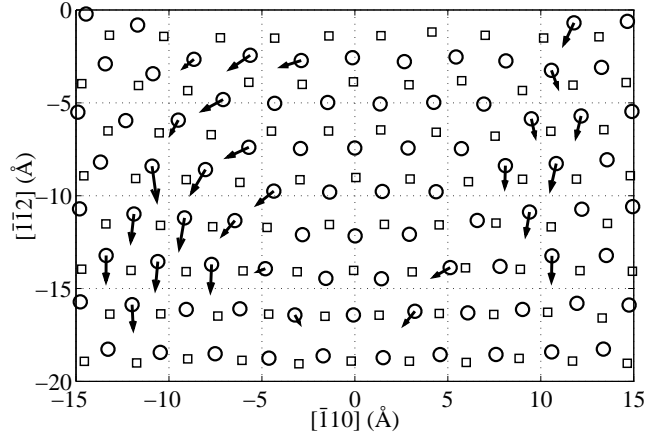}
	\caption{\label{fig:lmap} Slip vector $\bm{s}_{12}$ of the
atoms on $(111)$ plane. \ccirc{1}~represent the atoms of the slipped
plane and $\square$ represent the atoms of the unslipped plane
adjacent to the slipped region. The slip is along the direction in
which the resolved shear stress has attained the critical value.}   
\end{minipage}
\end{figure}

\begin{figure*}
\begin{minipage}[ht]{7in}
	\ig{7}{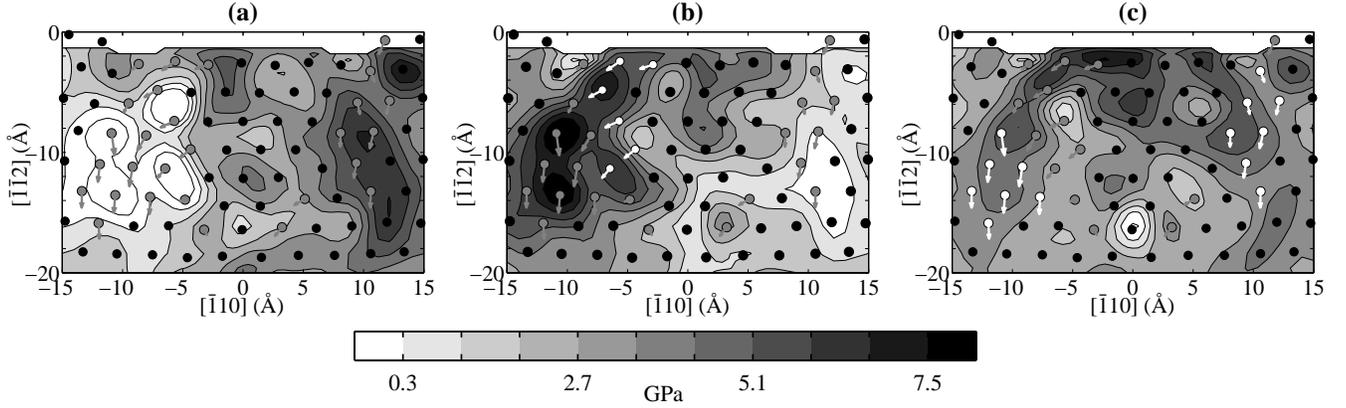}

	\caption{\label{fig:rss2} Resolved shear stresses on $(111)$
plane, at the second yield point, along the direction: (a)
$[12\bar{1}]$; (b) $[21\bar{1}]$; (c) $[11\bar{2}]$. \ccirc{1}~represent
the atoms with slip vectors coinciding with the RSS direction;
\ccirc{0.75} represent other slipped atoms and \ccirc{0}~represent
the atoms in the undeformed region. The observed slip is
not necessarily in direction of the maximum resolved shear stress.}
\end{minipage}
\end{figure*}

Further indentation results in a second yield, at
which point dislocation loops are nucleated on the slip planes outside
the defect as shown in Fig.~\ref{fig:lock}. Figure~\ref{fig:lmap} shows
the slip vectors $\bm{s}_{12}$, of the atoms on the $(111)$ plane
dislocated during the second yield and the dislocation loops extending
beyond the stair-rods can be seen. The contour plots of RSS on $(111)
\langle112\rangle$ along with the slip vectors are shown in
Fig.~\ref{fig:rss2}. It is seen that the activated slip direction is
not along the maximum RSS direction, but along the direction in which
the RSS has reached the critical value. This observation further
corroborates the discussion presented above. Thus, deformation results
through a sequence of elastic and plastic responses, with the
elastic responses culminating in plastic events.

\subsection{Retraction}

When the indenter is retracted after the first yield point, the force
curve retraces the indentation path at small displacements. This
suggests that the defect nucleated at the first yield point has
disappeared and the substrate has recovered is original undeformed
state upon retraction. However, the force curve during retraction
after the second yield point signifies permanent deformation.

Upon retracting the indenter,
the external stress vanishes and the internal compressive stress
dominates the dynamics of deformation. It is seen above that the
interior of the pyramidal defect is under enormous compressive stress
due to the strain imposed during the first yield. This compressive
stress when resolved onto the slip planes is opposite in
direction to the RSS during indentation. A restoring force is induced
that is strong enough to effect the unzipping of the stair-rods into
their constituent partials, which then glide toward the surface
healing the stacking fault along the way. Eventually, the defect
disappears and the substrate recovers its original configuration with
no residual deformation as seen in Fig.~\ref{fig:F-z}.

However, after the second yield, the dislocation loops extend
beyond the stair-rods leading to 
dislocation locking. And upon retracting the indenter after the second
yield, the aforementioned restoring forces are not strong enough to
unlock the locked structure, thus giving rise to permanent plastic
deformation observed in Fig.~\ref{fig:F-z}.

\section{Concluding Remarks}

We have investigated the atomistic mechanisms of plastic deformation
during nanoindentation of Au (001) surface with a noninteracting
indenter. A recently developed slip vector analysis has been employed
to identify the defect structures formed during initial plastic
yield. During indentation, the accumulated elastic energy in the
indented region is partially relieved by the nucleation of a
pyramidal defect structure. The defect is formed by the surface
nucleation of Shockley partials on the four $\{111\}$ slip planes at the
periphery of the contact region. These partials glide away from the surface
creating stacking faults that grow in size and intersect with those
on the adjacent planes. At the intersections, the partials zip to form
sessile stair-rods which contribute to the strain hardening observed
after the first yield. The observed slip is in the most energetically
favorable direction, which corresponds to the direction in which the
RSS has reached the critical value and is not necessarily the maximum
value. The CRSS estimated in this study is in the range of $1.8-2.3$
GPa in excellent agreement with the experimental estimates. 

Upon retracting the indenter after the first yield, the pressure
due to the compressive strain in the defect induces restoring forces
that heal the plastic deformation. Further indentation results in a
second yield that causes the
dislocation loops to extend beyond the stair-rods forming dislocation
locks. The unlocking forces of these structures is greater than the
internal restoring forces active during indenter retraction and thus
effect permanent deformation after the second yield.

We proposed a three step mechanism based on dislocation theory that
elucidates the physics behind the formation of the observed defect
structures during gold nanoindentation. According to this mechanism,
the defects produced depend on the crystallography of the indented
surface as seen in experiments.  

\section{Acknowledgments}
This work was supported by NSF Career Grant BES-9983735. The
calculations were performed using parallel computing resources of High
Performance Computing Partnership at Iowa State University.

\bibliography{paper}

\begin{thebibliography}{35}
\bibitem[{sch()}]{schmidfac}
\bibinfo{note}{The Schmid factors for $\bm{\sigma} = \sigma_{zz}$ on $(111)$
  along $[10\bar{1}]$, $[01\bar{1}]$ and $[11\bar{2}]$ are 0.4082, 0.4082 and
  0.4714 respectively.}
\end{thebibliography}

\end{document}